\documentclass[letterpaper,12pt]{article}   
\usepackage{osajnl}
\usepackage{breqn}

\begin{document}

\title{Multiple Isotope Magneto Optical Trap from a single diode laser}

\author{V.M. Valenzuela, S. Hamzeloui, M. Guti{\'{e}}rrez and E. Gomez$^*$}
\address{Instituto de F{\'{i}}sica, Universidad Aut{\'{o}}noma de San Luis Potos{\'{i}}, San Luis Potos{\'{i}} 78290, M{\'{e}}xico}
\address{$^*$Corresponding author: egomez@ifisica.uaslp.mx}

\begin{abstract}We present a Dual Isotope Magneto Optical Trap produced using a single diode laser. We generate all the optical frequencies needed for trapping both species using a fiber intensity modulator. All the optical frequencies are amplified simultaneously using a tapered amplifier. The independent control of each frequency is on the RF side rather than on the optical side. This introduces an enormous simplification for laser cooling applications that often require an acousto-optic modulator for each laser beam. Frequency changing capabilities are limited by the modulator bandwidth (10 GHz). Traps for more isotopes can be simply added by including additional RF frequencies to the modulator.
\end{abstract}


\maketitle 

\section{\label{introduction}Introduction}

Magneto Optical Traps (MOT) are the starting point for many experiments in atomic physics. Alkali atom MOTs are quite common and they require the use of two lasers, i.e., the trap and the repumper lasers. There is often the desire to trap different species simultaneously. This is the case, for example, for the study of quantum behavior of mixtures of bosons and fermions [1,2], multiple species photo-association [3,4], Feshbach resonances between different atoms [5], sympathetic cooling [6], or comparing properties of different atoms [7] among others. The additional infrastructure required for having more species often limits its use.

We present a system capable of producing a Multiple Isotope Magneto Optical Trap (MIMOT) using a single diode laser. The system uses a fiber intensity modulator to generate sidebands at the required optical frequencies and a tapered amplifier to boost the power. We demonstrate the system by trapping $^{85}$Rb and $^{87}$Rb simultaneously, but additional isotopes could be added if the appropriate RF frequencies are injected into the fiber modulator. The system may also be useful in situations where multiple beams are necessary, {\it e.g.}, for trapping atoms with electronic structure more complicated than alkali atoms [8], for laser cooling of molecules [9] and  for the generation of many frequencies for white light cooling [10].

Dual Isotope Magneto Optical Traps (DIMOT) have been demonstrated for Rb [2,4,6,7,11-14], K [15,16], Yb [17], He [18], Li [19,20], Sr [21,22] and Cr [23]. For some elements the isotope shift is small enough that the frequency difference can be covered with acousto-optic modulators (AOM) [1,15-17,21]. For atoms with a larger isotope shift 4 independent lasers are often used, 2 for trapping and 2 for repumping [1,20]. Each laser beam needs to be frequency locked and requires its own AOM to control the frequency and power.

A trick to reduce the number of lasers is to add light modulation to produce sidebands and use them for the extra optical frequencies needed. One implementation uses current modulation in a slave laser to derive both the trap and repumper out of it [24]. The sideband plays the role of the repumper due to the low power available. Electro-optical modulators (EOM) are another common implementation [13,14,25]. EOMs are simpler to use than current modulation and they give good modulation efficiency at high RF frequencies, but they are usually resonant devices that only work around certain frequency. Fiber modulators have all the benefits of an EOM and a very large bandwidth (10 GHz). Their only problem is that they can only handle low optical power and an optical amplifier is needed to boost the power.  

The system we present combines the very large bandwidth of a fiber modulator with the high optical power of a tapered amplifier. The use of a double pass configuration for the tapered amplifier was essential in going from 0.5 mW out of the fiber modulator to 1 W. The power is high enough that the sideband can now be used as the trapping beam and not only as the repumper as before. Due to the large bandwidth we include different RF frequencies using the same device to obtain the DIMOT. The tuning range for any beam is two orders of magnitude larger than what can be achieved using AOMs and it is big enough (for a 40 GHz modulator) to cover even the He isotopic shift [18]. The independent control of all the beams is now on the RF side. There is no need to add extra AOMs to control the frequency and power of each beam. All the beams propagate in the same direction and it is not necessary to spatially overlap them as would be the case when using AOMs. A similar idea has been used to obtain the repumper light in a single isotope trap [26].

\section{\label{setup}Experimental setup}

The experimental setup is very simple (Fig.~1). We lock a diode laser to a saturation spectroscopy reference. We send the laser light into a fiber intensity modulator (EOSPACE AZ-0K5-10-PFU-SFU-780) that generates multiple sidebands at the optical frequencies needed for trapping. The modulator handles only low optical power (below 10 mW) and we amplify the output light using a double pass tapered amplifier (Eagleyard EYP-TPA-07800-01000) [27,28]. The amplifier saturates at 0.5 mW of input light to get 1 W out. The input saturation power is lower than reported before [27] probably because of better mode matching to the tapered amplifier from the output fiber of the modulator. The light then goes into the three retro-reflected MOT beams. No other beam is needed at this point, since the position of the sidebands match the required optical frequencies for trapping both isotopes. Traps with three different isotopes [1] could be easily implemented using this system. Since we lock the laser to one of the repumping transitions we need to add only three different RF frequencies to the modulator to get a DIMOT.

Figure~2 (upper trace) shows the Doppler broadened spectrum of rubidium. The two outer peaks correspond to the ground hyperfine splitting of $^{87}$Rb and the two inner ones to $^{85}$Rb. Each Doppler broadened peak contains three excited state hyperfine transitions indicated by black vertical lines (middle trace). We lock the laser to the 5S$_{1/2}$ F=1 to 5P$_{3/2}$ F=2 transition of $^{87}$Rb that corresponds to the repumping transition. The position of this (carrier) laser is indicated by the tallest line in the lowest trace of Fig.~2. We produce sidebands with a RF frequency of 6589 MHz for the $^{87}$Rb trap laser, 5458 MHz for the $^{85}$Rb trap laser and 2545 MHz for the $^{85}$Rb repumper laser (lower trace of Fig.~2). The trapping lasers are red detuned from resonance by 19 MHz. The $^{85}$Rb repumper sideband is smaller since the power needed is less than that for the trapping beams. The trap size is relatively constant over a broad repumper frequency range. All the blue detuned sidebands (Fig.~2) are far from resonance and they have no noticeable effect on the atoms.

We use the simplest implementation that gives us control over the repumper of $^{85}$Rb, but not over the repumper of $^{87}$Rb since that corresponds to the carrier. This last beam could be set out of resonance with the use of AOMs to shift the frequency, or by locking the laser at a different frequency. An extra RF frequency must be injected into the modulator to recover and control the repumper beam of $^{87}$Rb. Turning the light completely off in our implementation must be done after the tapered amplifier. If an AOM is used to do it then that will introduce an extra frequency shift for all the beams that needs to be taken into account when calculating the fiber modulator RF frequencies. An alternative is the use of very fast mechanical shutters that introduce no frequency shift [29].

The fiber modulator gives all the flexibility needed for atomic physics experiments. For example, to do optical molasses one needs to rapidly change the detuning of the trap laser. This is achieved by simply changing the corresponding RF frequency. We turn off a particular beam by just removing the RF, something that is done fast and with high extinction ratio. An additional beam for optical pumping to the lower ground hyperfine level can be obtained by adding the corresponding RF frequency to the fiber modulator.

\section{\label{rfcontrol}RF control}

The RF system for the fiber modulator is shown in blue dashed lines in Fig.~1. We have three separated oscillators to generate the required RF frequencies. Each oscillator is followed by a RF switch (Minicircuits ZFSWA2-63DR+) to turn off that frequency. We combine the different frequencies using a RF splitter/combiner (RF Lambda RFLT3W2G08G) that has low insertion loss (1 dB) and good isolation (18 dB). We amplify the RF signals with a 35 dB amplifier (Minicircuits ZVE-3W-83+) that covers the entire range and we send this signal to the fiber modulator.

There are two options for the oscillators, one based on PLL synthesizers that are very stable and the other one based on voltage controlled oscillators (VCO) that are simpler to use. For the first option we produce the 2545 MHz signal directly from a PLL synthesizer (EVAL-ADF4350-EB2Z) that has a precision better than 1 Hz. The 5458 MHz signal uses the same synthesizer model but we double (Minicircuits ZX90-2-36S+) and filter (Minicircuits VBFZ-5500-S+) the RF frequency to eliminate any unwanted harmonics. The same scheme works for the 6589 MHz signal but we use instead a synthesizer (Gigatronics 1018) that directly gives the right RF frequency. This choice of oscillators is useful for situations when high precision in frequency is important, but it is not very practical for making frequency changes.

Alternatively, we have also used VCOs that are readily tunable with voltages derived from our control system [27]. The VCO frequency depends on voltage and temperature variations. We derive the VCO control voltage from a very stable voltage reference (LM399) and we change the set voltage slightly with the control system in the laboratory. The sensitivity to voltage is around 100 MHz/V and we measure frequency variations below 1 MHz even without temperature stabilization of the oscillator. A VCO with smaller voltage sensitivity gives better stability and using them we have observed frequency variations below 100 kHz. The stability achieved is more than enough for many applications in atomic physics including atom trapping. The VCO is very convenient to introduce fast frequency changes. It can sweep at a rate of 150 MHz/$\mu$s, and the optical amplifier reduces that to 30 MHz/$\mu$s which is more than enough for typical applications in laser cooling.

The RF switch has a power reduction of 40 dB in the off mode. It is capable of turning off the laser beam in just 35 ns (3 ns switches are available). This is very convenient not only for atom trapping but also for measurements that require fast turn off times. For example, lifetime measurements require turn off times of less than 10 ns, which is usually achieved using Pockels cells and AOMs [30]. Pockels cells have fast switching times (a few ns) but a bad extinction ratio (less than 20 dB) and they have to be shielded since they broadcasts a lot of RF; AOMs have good extinction ratio (40 dB) but they are slower (more than 30 ns). There is an advantage in turning off an RF signal rather than a DC signal. Using the fiber modulator one gets fast switching (3 ns), good extinction ratio (40 dB) and almost no RF broadcasted. A fiber modulated beam could be added to a traditional trapping setup, but here it is already included in the system.

We use a single broadband RF amplifier for all the frequencies. We inject low RF powers (3.2 dBm at 6589 MHz, 3 dBm at 5458 MHz and -3.2 dBm at 2545 MHz) into the fiber modulator to reduce nonlinear effects as we explain in the next section. The nonlinearity at the RF amplifier introduces two extra frequency components equal to the sum and difference of the trapping beams RF frequencies. The amplitude of these nonlinear components however are 27 dB below the trapping RF frequency level. Any spurious frequency becomes important only if it happens to be on resonance with some transition. The spurious frequency that gets closest to a particular resonance is 312 MHz away and it is not visible up to a power 63 dB below the trap RF power. 

\section{\label{opticalpart}Optical modulation and amplification}

The fiber modulator is designed to be used in the TM mode with a voltage for a $\pi$ phase shift in the modulator of $V_\pi=1.1$ V at DC. It has an insertion loss of 3.2 dB and a DC extinction ratio of 19 dB. The input light polarization has to be correct to better than a degree for the modulator to behave stably, otherwise even a change of 0.2 $^o$C is enough to change the polarization to the perpendicular mode. We optimize the input polarization but also control the temperature to 0.2 $^o$C and we get very stable operation. An additional slow feedback system keeps the output power of the fiber modulator at one half of the maximum output power to have the best sideband efficiency. The output power lock is not necessary when using a phase modulator instead of an intensity modulator but the nonlinear response will change as well.

The modulator changes the phase of one of the arms of a Mach-Zehnder interferometer with the applied voltage in order to change the intensity [31]. The electric field at the output is given by

\begin{equation}
E=\frac{E_0}{2} \left[ \cos{(wt)} + \cos{\left( wt+\pi \frac{V_0}{V_{\pi}} \right)} \right], \label{electricmod}
\end{equation}
with $w$ the light frequency and $V_0$ the input voltage to the modulator. Figure 3 (red squares) shows the first order sideband amplitude as a function of the input RF power when we modulate at 5.6 GHz. The input RF power should not exceed 27 dBm for our modulator model. We measure the sideband amplitude by looking at the transmitted light on a Fabry-Perot cavity to identify each optical frequency separately. The solid red line is a fit to the theoretical dependence on RF power given by

\begin{equation}
S_1=J_1^2 \left( \frac{\pi V_m}{V_{\pi}} \right), \label{sideband1}
\end{equation}
where $V_m$ is the amplitude of the modulating RF voltage and $J_1$ is the Bessel function of the first kind. The Appendix gives the derivation of Eqs. 2 to 4. The fit to the data gives a $V_{\pi}=3.32 \pm 0.08$ at 2.7 GHz, $V_{\pi}=4.17 \pm 0.08$ at 5.6 GHz and $V_{\pi}=3.80 \pm 0.06$ at 6.8 GHz. At high input RF powers the modulator generates a second order sideband. The size of the second order sideband is shown in Fig.~3 (green triangles) together with the theoretical value (solid green line) given by

\begin{equation}
S_2=J_2^2 \left( \frac{\pi V_m}{V_{\pi}} \right). \label{sideband2}
\end{equation}

The fiber modulator output power ($P$) is set by the DC voltage port. The visibility $\upsilon=(P_{max}-P_{min})/(P_{max}+P_{min})$ of the modulator output power decreases as we increase the input RF power (blue diamonds in Fig.~3). The visibility is reduced since the photodetector averages the modulated signal over a certain voltage range. The solid blue line is the theoretical value given by

\begin{equation}
S_L=J_0 \left( \frac{\pi V_m}{V_{\pi}} \right). \label{sidebandv}
\end{equation}
A good visibility is important in order to keep the output power of the laser locked. The visibility goes to zero at a particular input RF power and the output intensity of the fiber cannot be locked there. The visibility becomes negative (reverses sign) at even higher RF powers and one must invert the sign of the feedback in order to keep the output power locked.

All the theoretical curves in Fig.~3 are obtained with a single set of parameters determined by a fit to only the first order sideband data. Even when better results could be achieved with independent fits, the agreement obtained using a single set of parameters shows a clear understanding of the process. The small discrepancy observed in the visibility fit may be due to the finite bandwidth of the detector used.

The first order sideband amplitude is maximum at a power of 18 dBm (corresponding to about 30\% of the total optical power in a single sideband), but at this power one gets a considerable amount of second order sideband. Also the modulation intensity visibility goes to zero and it is not possible to lock the modulator output power. Adding more than one RF frequency produces quite a bit of frequency mixing at high RF powers. We operate at lower RF powers (3 dBm) to reduce the appearance of additional optical frequencies and we have sidebands with about 7\% of the total optical power.

Figure 4 shows the optical spectrum after the tapered amplifier obtained using a Fabry-Perot cavity with a 1.5 GHz free spectral range. The sidebands corresponding to the traps ($\nu_1$= 6589 MHz and $\nu_2$= 5458 MHz) and repumper ($\nu_3$= 2545 MHz) optical frequencies have a height equal to 7$\%$, 8$\%$ and 4$\%$ of the carrier. Additional peaks corresponding to the sum (12047 MHz) and difference (1130 MHz) of the trapping RF frequencies are also visible but they are very far from any resonance to be of importance. Blue detuned sidebands are also visible and make the figure symmetric.

The input RF power used (3 dBm) gives good modulation efficiency but still lets us treat each trapping optical frequency almost independently since changes of one introduce variations on the other sideband smaller than 10\%. This is not the case at higher RF powers where we see considerable frequency mixing (additional optical frequencies present) and a strong influence on the presence or absence of one sideband over the size of the other. At low RF levels the modulator output power remains stably locked when switching the RF on and off.

The tapered amplifier gives high output power (1 W) and it is simpler to use than diode laser injection locking [12]. It has a gain range of 20 nm around 780 nm which is more than enough for isotope shifts, but it is usually not enough for different atomic elements. We observed no additional optical frequency mixing introduced by the tapered amplifier. Frequency mixing is expected to be small since any frequency difference is larger than 1 GHz [32]. Thermal fluctuations in the amplifier change the power distribution between carrier and sidebands. This effect seems to come from an etalon effect due to reflections on the end faces of the tapered amplifier. The effect is more evident for us probably due to the competition of amplification at different optical frequencies. At the highest current (3 A) in the tapered amplifier we see fluctuations of 20\% on the sideband efficiency for a period of 5 minutes whereas at lower currents (2 A), where the heat load is smaller, the fluctuations are reduced to 7\% for the same time scale. The fluctuations should improve with better temperature stability. Both the amplifier and fiber modulator introduce no reduction on the speed of light switching.

The input optical power to the tapered amplifier must not change rapidly to avoid the risk of damage. Feed forward strategies are applied when the amplifier is preceded by AOMs [27]. When using a fiber modulator this is not a big concern since the modulator output power remains constant as we change either the frequency or power of the RF signal injected into the modulator. Switching the RF off suddenly transfers the sideband power to the carrier keeping the total laser power constant.

\section{\label{trappingresults}Trapping results}

We use the system to obtain a simultaneous trap of the two stable rubidium isotopes. We typically capture 4 $\times$ 10$^8$ atoms of $^{87}$Rb and a similar amount of $^{85}$Rb on steady-state. The atom number of each isotope depends on the laser power of the trapping beams. After some beam shaping of the tapered amplifier light, we send 260 mW of optical power to the atoms. Each trapping beam has a diameter of 3 cm and an intensity of 7.6 mW/cm$^2$. The sideband corresponding to $^{87}$Rb ($^{85}$Rb) has 7\% (8\%) of the power, and therefore the trapping intensity is 0.5 (0.6) mW/cm$^2$  respectively. 

Figure 5 shows the number of atoms trapped as a function of time. At t=0 all RF switches are off and there are no atoms trapped. The sequence is: $^{87}$Rb only - $^{87}$Rb and $^{85}$Rb - $^{87}$Rb only, with each stage lasting for 10 s and we turn both traps off at the end. A clear increase is observed when both traps are on. Also the fluorescence does not go to zero at t=20 s since one of the isotopes still remains trapped. We imaged both traps separately to further verify that both isotopes are trapped simultaneously and we see a good overlap between the two clouds. The trap has a lifetime of 6 s limited by the quality of the vacuum. The noise observed in Fig.~5 is partially due to the stability of our frequency locking system, but it has also some contribution from the intensity fluctuations of the sidebands described before. The steady state atom number remains constant for longer times up to fluctuations like those shown in Fig. 5.

The two trapping beams are fairly independent of each other since turning one beam off changes the power on the other beam by no more than 10\%. Working at high RF power gives a nonlinear response but on the other hand it helps increasing the number of atoms. Setting the RF power of $^{87}$Rb (6589 MHz) close to the diffraction efficiency maximum of Fig.~3 translates into a 7 fold increase in trapping power and gives a trap 3 times bigger. This means that we can trap in excess of 10$^9$ atoms.

\section{\label{trapcar} Conclusion}

We present a system capable of trapping multiple isotopes simultaneously in a MOT. The system uses a single diode laser and a fiber intensity modulator to produce all the required optical frequencies. They are amplified simultaneously with a tapered amplifier to obtain high output power. The control of the different optical frequencies and their intensities is now moved to the RF side. This gives a considerable simplification over traditional setups. As an example, a typical setup for a Dual Isotope Magneto Optical Trap would require four lasers with their corresponding locking systems, four acousto-optic modulators and many optical components. This is replaced here by a single laser and a fiber modulator. Everything else is controlled electronically. We believe that this setup will promote the use of multiple species in atomic physics experiments.

\section*{Acknowledgments}

We acknowledge support from CONACYT, Promep and UASLP. We thank Jonathan Espinosa for his assistance on electronics.

\section*{Apendix}

Here we derive Eqs.~2 and 3 for the diffraction efficiency. Start with Eq.~1 with a time dependent voltage

\begin{equation}
V_0=V_{DC}+\beta \sin{(\Omega t)}, \label{V0vary}
\end{equation}
where $V_{DC}$ is the voltage on the DC port of the fiber that is locked at half power, $\Omega$ is the modulation frequency and $\beta$ the modulation amplitude. The resulting function is expanded in terms of Bessel functions of the first kind

\begin{dmath}
E = \frac{E_0}{2} \left\{ \cos{(wt)} +\cos{ \left( wt+ \pi \frac{V_{DC}}{V_{\pi}} \right) } \left[ J_0\left( \pi \frac{\beta}{V_{\pi}} \right) + 2 \sum^\infty_{n=1} J_{2n} \left( \pi \frac{\beta}{V_{\pi}} \right) \cos{(2n\Omega t)} \right] - \sin{ \left( wt+ \pi \frac{V_{DC}}{V_{\pi}} \right) } \left[ 2 \sum^\infty_{n=1} J_{2n-1} \left( \pi \frac{\beta}{V_{\pi}} \right) \sin{((2n-1)\Omega t)} \right] \right\}.
\end{dmath}

From the above expression one recognizes the carrier and sideband amplitudes. The modulus square of the Fourier transform of this equation gives the spectrum and results in Eqs.~2 and 3. Equation~4 comes from the modulus square of Eq.~6 that gives the intensity, but since the detector is slow one must take the average over a RF period.

\section{References}

\bibliographystyle{osajnl}


\clearpage

\section*{List of Figure Captions}

Fig. 1. Experimental setup. In solid red lines are the optical paths and in dashed blue lines the RF parts.

\noindent Fig. 2. Spectrum of rubidium. The upper trace (dashed black line) shows the Doppler broadened spectrum of rubidium. The middle trace (solid black line) shows the excited state hyperfine transitions with the zero frequency corresponding to the left most transition. The lower trace (solid red line) shows the spectrum of the laser light out of the fiber modulator. The tallest peak is the carrier and the rest are sidebands. The three left sidebands are tuned to the required transitions for the DIMOT. The laser is locked at the carrier position. The energy levels for the two isotopes are shown on the right.

\noindent Fig. 3. Fiber modulator characterization. It shows the first order sideband amplitude (red squares), second order sideband amplitude (green triangles) and visibility of the DC locking signal (blue diamonds). The solid lines correspond to the theoretical values. The fit is done to the first order sideband to obtain the three curves as indicated in the text. The second order sideband signal is buried in the noise for RF powers below 13 dBm. 

\noindent Fig. 4. Optical spectrum after the tapered amplifier measured in a Fabry-Perot cavity. The tallest peaks indicate the carrier and free spectral range of the cavity (1.5 GHz). The peaks with $\nu_1$=6589 MHz and $\nu_2$=5458 MHz correspond to the trapping frequencies and the one with $\nu_3$=2545 MHz is the repumper frequency. The nonlinearities and blue detuned sidebands are also visible.

\noindent Fig. 5. Atom number in the DIMOT. The sequence is: $^{87}$Rb only - $^{87}$Rb and $^{85}$Rb - $^{87}$Rb only, with each stage lasting for 10 s and we turn traps off at the end.

\clearpage

\begin{figure}[htbp]
\centering
\includegraphics[width=8.3cm]{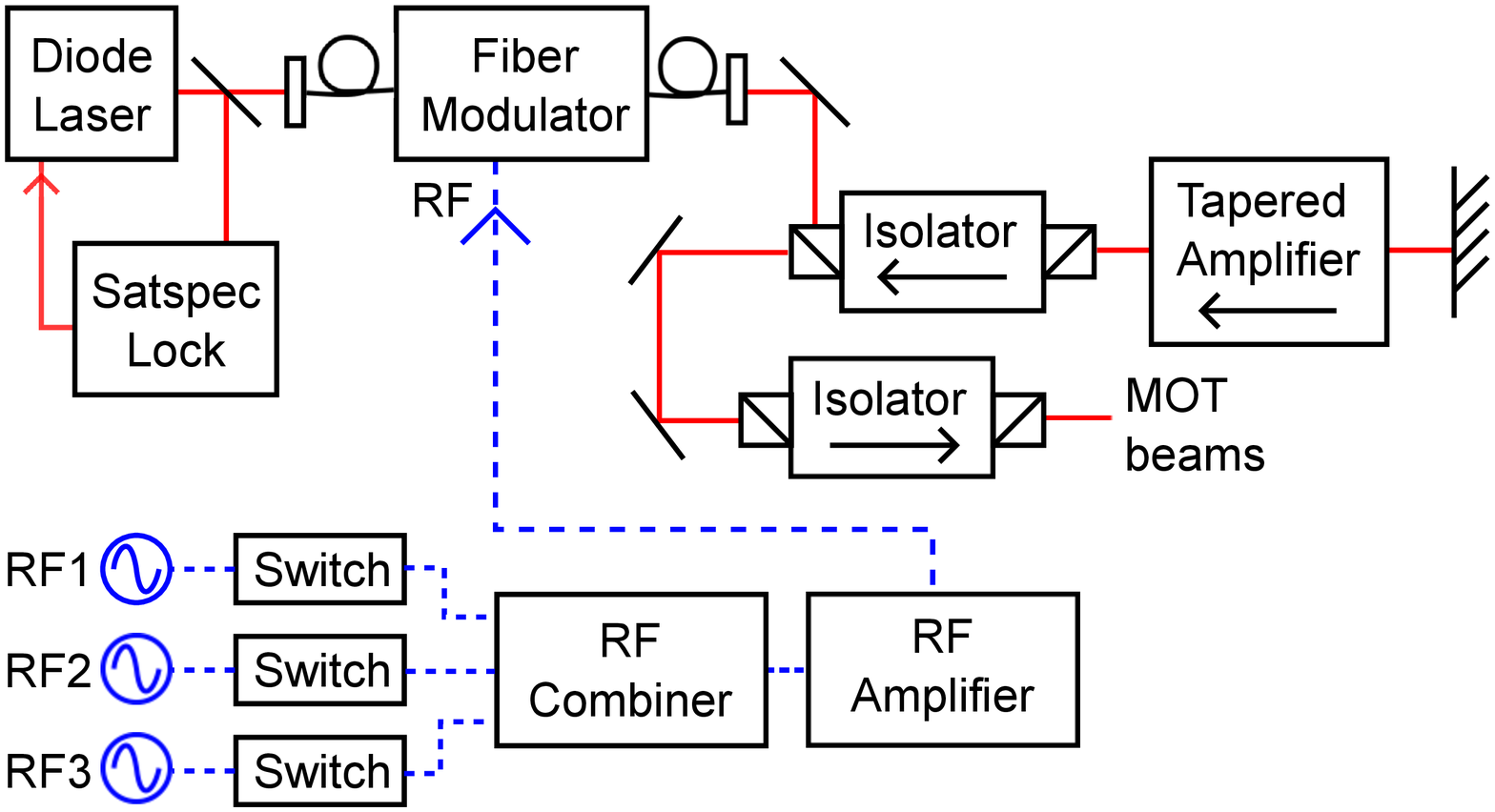}
\caption{\label{expsetup} Experimental setup. In solid red lines are the optical paths and in dashed blue lines the RF parts.}
\end{figure}

\clearpage

\begin{figure}[htbp]
\centering
\includegraphics[width=8.3cm]{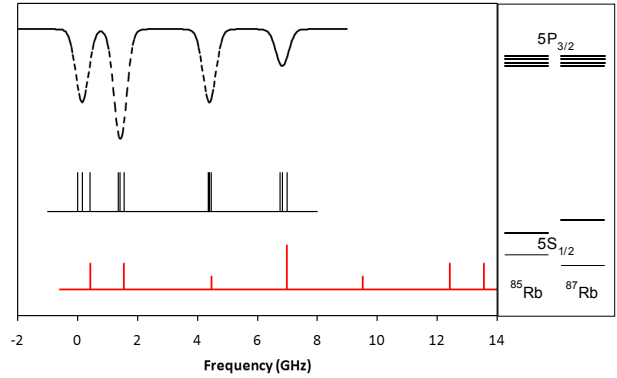}
\caption{\label{spectrumrb} Spectrum of rubidium. The upper trace (dashed black line) shows the Doppler broadened spectrum of rubidium. The middle trace (solid black line) shows the excited state hyperfine transitions with the zero frequency corresponding to the left most transition. The lower trace (solid red line) shows the spectrum of the laser light out of the fiber modulator. The tallest peak is the carrier and the rest are sidebands. The three left sidebands are tuned to the required transitions for the DIMOT. The laser is locked at the carrier position. The energy levels for the two isotopes are shown on the right.}
\end{figure}

\clearpage

\begin{figure}[htbp]
\centering
\includegraphics[width=8.3cm]{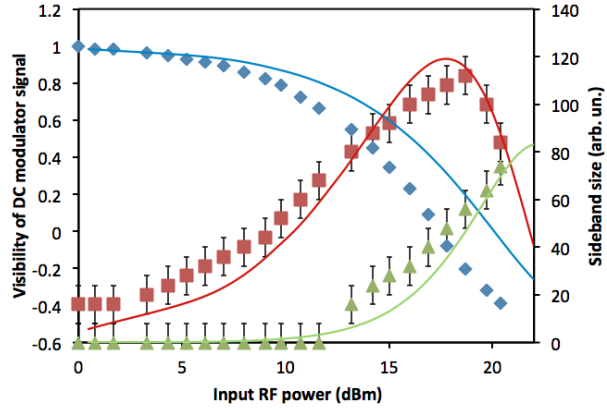}
\caption{\label{dcmodulation} Fiber modulator characterization. It shows the first order sideband amplitude (red squares), second order sideband amplitude (green triangles) and visibility of the DC locking signal (blue diamonds). The solid lines correspond to the theoretical values. The fit is done to the first order sideband to obtain the three curves as indicated in the text. The second order sideband signal is buried in the noise for RF powers below 13 dBm.}
\end{figure}

\clearpage

\begin{figure}[htbp]
\centering
\includegraphics[width=8.3cm]{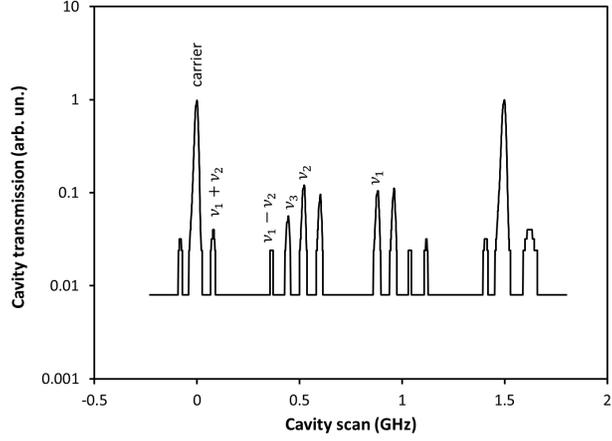}
\caption{\label{fabrypeaks} Optical spectrum after the tapered amplifier measured in a Fabry-Perot cavity. The tallest peaks indicate the carrier and free spectral range of the cavity (1.5 GHz). The peaks with $\nu_1$=6589 MHz and $\nu_2$=5458 MHz correspond to the trapping frequencies and the one with $\nu_3$=2545 MHz is the repumper frequency. The nonlinearities and blue detuned sidebands are also visible.}
\end{figure}

\clearpage

\begin{figure}[htbp]
\centering
\includegraphics[width=8.3cm]{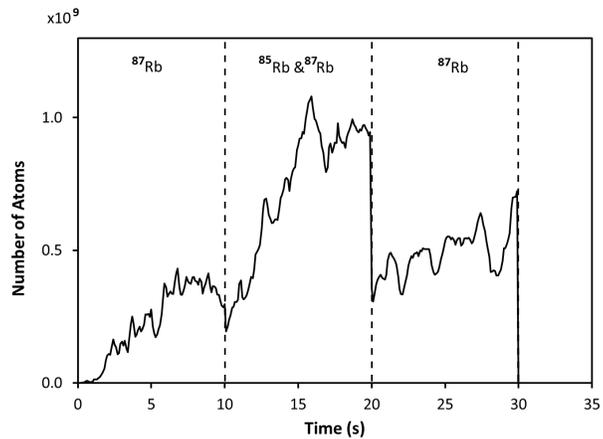}
\caption{\label{atomnum} Atom number in the DIMOT. The sequence is: $^{87}$Rb only - $^{87}$Rb and $^{85}$Rb - $^{87}$Rb only, with each stage lasting for 10 s and we turn traps off at the end.}
\end{figure}

\end{document}